\documentclass[aps,pra,floatfix,superscriptaddress,showpacs,twocolumn]{revtex4}
\usepackage{amsmath}
\usepackage{graphicx}
\usepackage{psfig}
\newcommand{\prn}[1]{\left(#1\right)}

\begin{document}
\title{Nonlinear magneto-optical rotation, Zeeman and hyperfine relaxation of potassium atoms in a paraffin-coated cell}
\author{J. S. Guzman}
\affiliation{Department of Physics, University of California,
Berkeley, CA 94720-7300}
\author{A. Wojciechowski}
\affiliation{Centrum Bada\'{n} Magnetooptycznych, Instytut Fizyki
im. M. Smoluchowskiego, Uniwersytet Jagiello\'{n}ski, Reymonta 4,
30-059 Krak\'{o}w, Poland}
\author{J. E. Stalnaker}\altaffiliation[Current Address: ]{National
Institute of Standards and Technology, 325 S. Broadway Boulder, CO
80305-3322} \affiliation{Department of Physics, University of
California, Berkeley, CA 94720-7300}
\author{K. Tsigutkin}
\affiliation{Department of Physics, University of California at
Berkeley, Berkeley, California 94720-7300}
\author{V. V. Yashchuk}
\affiliation{Advanced Light Source Division, Lawrence Berkeley
National Laboratory, Berkeley CA 94720}
\author{D. Budker}
\email{budker@berkeley.edu} \affiliation{Department of Physics,
University of California, Berkeley, CA 94720-7300}
\affiliation{Nuclear Science Division, Lawrence Berkeley National
Laboratory, Berkeley CA 94720}

\date{\today}
\begin{abstract}
Nonlinear magneto-optical Faraday rotation (NMOR) on the potassium
D1 and D2 lines was used to study Zeeman relaxation rates in an
antirelaxation paraffin-coated 3-cm diameter potassium vapor cell.
Intrinsic Zeeman relaxation rates of
$\gamma^{NMOR}/2\pi=2.0(6)\,{\rm Hz}$ were observed.  The relatively
small hyperfine intervals in potassium lead to significant
differences in NMOR in potassium compared to rubidium and cesium.
Using laser optical pumping, widths and frequency shifts were also
determined for transitions between ground-state hyperfine sublevels
of $^{39}$K atoms contained in the same paraffin-coated cell. The
intrinsic hyperfine relaxation rate of $\gamma^{hf}_{expt}/2\pi =
10.6(7)\ $Hz and a shift of $-9.1(2)\ $Hz were observed. These
results show that adiabatic relaxation gives only a small
contribution to the overall hyperfine relaxation in the case of
potassium, and the relaxation is dominated by other mechanisms
similar to those observed in previous studies with rubidium.
\end{abstract}
\pacs{33.55.Ad,33.55.Fi,32.30.Bv,32.70.Jz,32.80.Bx,95.55.Sh}

\maketitle
\section{Introduction}
The drive to develop small atomic clocks, gyroscopes, and
magnetometers has drawn interest to miniature antirelaxation-coated
alkali-atom vapor cells near $1\,{\rm mm}^3$ in volume (see for
example Ref.\ \cite{Bud2005NIST} and references therein). The large
surface-to-volume ratio of small vapor cells means that the
interactions of the confined atoms with the cell walls are
particularly important. These interactions can cause spin
de-coherence and limit the performance of such cells in atomic
clocks and magnetometers. The use of antirelaxation coatings reduces
these effects and may provide an alternative to small buffer gas
cells \cite{Bal2006}.

In this research, we study nonlinear magneto-optical rotation and
the Zeeman and hyperfine relaxation rates in potassium atoms in a
paraffin-coated vapor cell. Similar investigations in rubidium and
cesium vapor cells have yielded information on relaxation mechanisms
for these atoms in coated cells \cite{Bud2005NIST,Gra2005}, and
contributed to the development of sensitive
nonlinear-optical-rotation magnetometers that are used, for example,
for detecting nuclear magnetization \cite{Yas2004,Xu2006IMAG}.

In addition to yielding information on the wall-relaxation
processes, these measurements are also motivated by an idea that the
use of K atoms in secondary frequency references (clocks) may be
advantageous compared to the use of Rb or Cs.

A figure of merit for a frequency reference is $\delta \nu/\nu$,
where $\nu$ is the frequency of the transition, and $\delta \nu$ is
the absolute frequency stability that scales with the transition
width.
Suppose that the width of the hyperfine-structure transition is
dominated by adiabatic collisions. It has been established that for
such collisions, the average phase shift during a collision is
proportional, among other factors, to the hyperfine-structure
frequency $\nu$  (see, for example, Refs. \cite{Van74,Bud2005NIST}
and a detailed discussion and further references in Ref.
\cite{VanierAudoin}, Ch. 3.4). Since the relaxation rate scales as
the square of the phase shift, a lower-frequency alkali
hyperfine-structure transition (for example, the hyperfine-structure
transition in $^{41}$K at $\nu\approx 254\ $MHz) may turn out to be
advantageous for a clock.

The significantly smaller hyperfine-structure intervals of potassium
lead to a considerably different, generally, much smaller nonlinear
Faraday rotation associated with creation and detection of atomic
alignment. Here we study some specific features of the nonlinear
magneto-optical rotation in potassium and measure the rate of its
spin relaxation. A comparison with the results for other alkali
atoms will enable a better understanding of the mechanisms that
cause spin relaxation, and aid in determining the feasibility of
making ultra-sensitive magnetometers and small atomic clocks using
paraffin-coated cells. This work is also complementary to the
studies of spin-exchange relaxation in potassium
\cite{Ale99_K,Ale99_K_ERRATUM}.


The cell studied in this work contains potassium with a natural
abundance
($^{39}\rm{K}\!:93.26\%,^{40}\rm{K}\!:0.012\%,^{41}\rm{K}\!:6.7\%$).
The overwhelming abundance of $^{39}\rm{K}$ in comparison with the
other two isotopes makes the contribution of this isotope to
magneto-optical rotation dominant. Thus in the following, we will
only be concerned with $^{39}\rm{K}$.

\begin{figure}[h]
\centerline{\psfig{figure=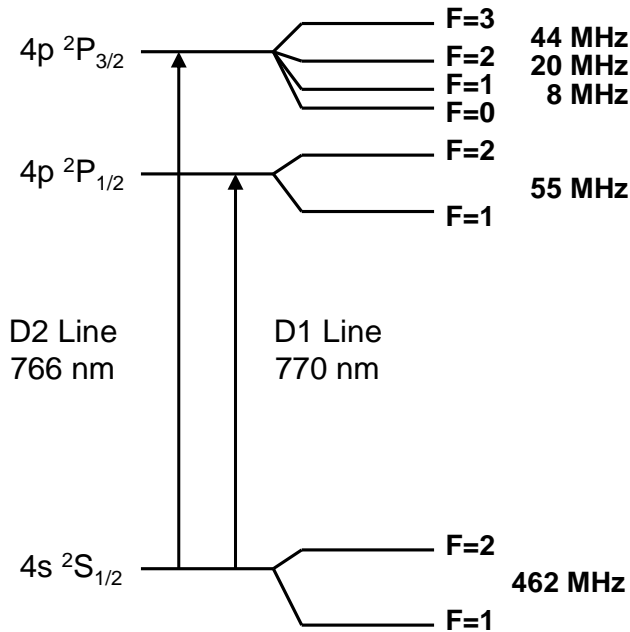,width=3.25in}}
\caption{Hyperfine splitting of low-lying levels of $^{39}{\rm K}$.}
\label{Kstructure}
\end{figure}

The hyperfine structure of $^{39}\rm{K}$ consists of $F=1,2$
hyperfine levels in the $4\,^{2}{\rm S}_{1/2}$ ground state, with an
energy splitting of $462\,{\rm MHz}$.  The upper levels of interest
in this work are the $4\,^{2}{\rm P}_{1/2}$
and $4\,^{2}{\rm P}_{3/2}$ (Fig.\ref{Kstructure}).

Faraday rotation due to coherence effects (see, for example, review
\cite{Bud2002RMP}) relies on the ability to create and detect atomic
alignment in the ground state. The alignment is produced via the
atoms' interactions with resonant linearly polarized light which
optically pumps the atoms. In the presence of a magnetic field, the
alignment evolves (in the case of weak light field, the evolution is
Larmor precession; a more complex evolution occurs at elevated light
powers \cite{Bud2000AOC}). The optical anisotropy of the evolved
medium results in a rotation of the light polarization.


The extent to which alignment can be created in the ground state by
optical pumping and the optical anisotropy associated with this
alignment both depend on the upper-state hyperfine structure being
resolved. The reason for this is that the electronic angular
momentum of the initial state is $J=1/2$, and such a state cannot
support polarization moments with rank $\kappa$ higher than $2J=1$.
Optical rotation via the mechanism discussed here requires
$\kappa=2$ (alignment; see, for instance, Ref. \cite{Bud2002RMP}).
In the limit where hyperfine structure is completely unresolved, we
can neglect the nuclear-spin part of the wavefunction. However,
since the electronic state cannot support alignment, it is
impossible to create or detect alignment of the initial state with
light.


In $^{39}\rm{K}$, the splitting of the $^{2}\rm{P}_{1/2}$ state is
55.6 MHz, and the splittings of the $^{2}\rm{P}_{3/2}$ state are
44.4 MHz, 20.4 MHz, and 7.5 MHz.  The Doppler width in this work is
$\approx 500\: \rm{MHz}$, thus there is a considerable suppression
of nonlinear Faraday rotation on both the D1 and D2 lines compared
with rubidium and cesium where the upper-state hyperfine intervals
are on the order of the Doppler width.
\section{Apparatus and Procedure}
We employ the technique of nonlinear magneto-optical rotation with
frequency modulated light (FM NMOR) \cite{Bud2002FM} and study the
zero-field FM NMOR resonance. The signal we detect is the amplitude
of optical rotation synchronous with the laser-frequency modulation.

\begin{figure}[h]
\centerline{\psfig{figure=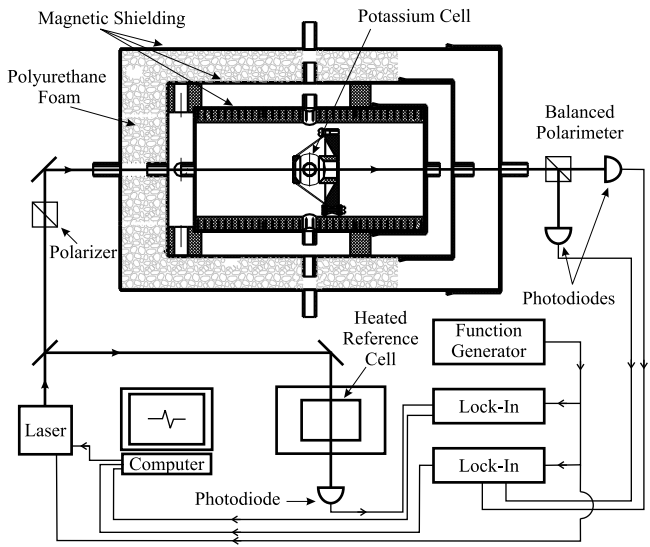,width=3.25in}}
\caption{Schematic diagram of experimental setup for measurement of
nonlinear magneto-optical rotation. For the measurement of the
transitions between hyperfine levels, the balanced polarimeter was
replaced with a single photodiode. An rf loop was introduced next to
the vapor cell trough one of the ports in the magnetic shield.}
\label{KApparatus}
\end{figure}

The experimental apparatus (Fig. \ref{KApparatus}) incorporates a
spherical glass cell with inner diameter of $3\ $cm and inner walls
coated with paraffin. A sample of solid potassium metal is contained
in a stem connected to the cell with an opening approximately $2\
$mm in diameter. The cell was made by M.~V.~Balabas using the
technology outlined, for example, in Ref. \cite{AleLIAD}.

The cell is placed within a magnetic shielding assembly. The
shielding consists of three nested 0.5-mm thick layers of Co-Netic
AA high-permeability alloy. Each layer is cylindrical in shape.
Starting with the innermost shielding layer, the radii of the layers
are 5.08, 8.26, and 11.43 cm, with lengths of 21.33, 26.90, and
35.56 cm, respectively. Within the magnetic shielding is a set of
coils for application of magnetic fields and gradients, allowing us
to reduce the effects of residual magnetization in the shielding,
and to investigate nonlinear magneto-optical rotation as a function
of the magnetic field. Magnetic fields can be applied along the
direction of the cylinder, which we define as the z-axis, as well as
along the two transverse axes, x and y. There are two additional
sets of coils for the $\partial{B_z}/\partial{z}$ and
$\partial{B_{x}}/\partial{x}-\partial{B_{y}}/\partial{y}$ gradient
fields.  The presence of magnetic fields transverse to the direction
of light propagation changes the shape of the longitudinal
magnetic-field dependence of the optical rotation \cite{Bud98}. By
applying magnetic fields in the transverse x- and y-directions, and
scanning the magnetic field in the z-direction, we were able to
compensate, using the observed lineshapes, the residual magnetic
fields along the x and y directions to within $\approx 0.5\:
\mu\rm{G}$.

Laser light from a New Focus Velocity diode laser (that could be
tuned to either D1 or D2 transition) was frequency modulated by
scanning the external cavity of the laser with a piezo-electric
actuator. The frequency of the light was modulated at 1 kHz with a
modulation depth of $\approx 400\: \rm{MHz}$. Part of the laser
light was sent to an uncoated potassium vapor cell used as a
reference for frequency stabilization purposes. The remaining light
was attenuated and sent to the coated vapor cell.

At room temperature the density of potassium atoms is $\approx\,3
\cdot 10^8\,{\rm atoms}/{\rm cm}^3$.  Heating the reference cell to
$\approx 100^{\circ}\rm{C}$ increases this density to $\approx\,5
\cdot 10^{11}\,{\rm atoms}/{\rm cm}^3$.
The laser light that passed through the reference cell was detected
with a photodiode and demodulated with a lock-in amplifier that
monitored the second harmonic of the photodiode signal. This signal
was digitized with a data acquisition board and was kept at a fixed
level by generating a computer-controlled voltage which was fed back
to the frequency-control input of the laser.

The coated vapor cell was also heated. Plastic tubing with 0.4-cm
diameter was wrapped into ten turns with a $\approx 10$-cm diameter.
These coils were attached to a Teflon cell holder using four Teflon
posts extending from the holder. Pressurized air was run through
copper tubing placed in boiling water and sent through the tubing
surrounding the cell. The hot air in the tubing convectively heated
the air surrounding the vapor cell. Based on unsaturated resonance
atomic absorption of $\approx 10 \%$ for the D1 transition and
$\approx 20 \%$ for the D2 transition, the atomic density was
estimated as $5\cdot 10^9\,{\rm cm^{-3}}$. This corresponds to an
estimated temperature of the cell of $47^{\rm o}$C (neglecting
possible deviations from the saturated-vapor density that may occur
in a coated cell).

The light that has passed through the vapor cell was analyzed with a
balanced polarimeter incorporating a polarization prism and two
large-area photodiodes. The output of the photodiodes was sent to
the differential input of a lock-in amplifier whose first-harmonic
output was digitized and recorded by the computer.

Nonlinear Faraday rotation was measured at fixed tuning of the
central frequency of the laser as a function of the longitudinal
magnetic field over the range of $\pm 800\,\rm{\mu G}$. Measurements
were done with laser-light power going into the coated cell between
2 and $115\,{\rm \mu W}$. Over this range of light powers, there was
considerable variation of the magnitude and magnetic-field
dependence of the Faraday-rotation.

The choice of the central laser frequency at which to lock was
determined by scanning over the potassium resonance with a fixed
magnetic field. The laser central frequency was then locked where
the Faraday rotation signal was the largest. Once the lock point was
determined, the magnetic field was scanned with decreasing and
increasing magnetic field values symmetrically around zero using a
computer-generated triangular wave.

Several acquisitions were averaged before changing the magnetic
field value. Scans at low light power, $<10\,{\rm \mu W}$ lasted
about 20 minutes, and scans at high light power, $>15\,{\rm \mu W}$,
lasted about 4 minutes.
Such long integration times were necessary
because the optical-rotation amplitudes at the lowest light powers
were as low as $10\ \mu$rad, while the noise of the polarimeter
exceeded the shot-noise limit
by a factor of a few.
\section{Results for Zeeman Relaxation Rates}
\begin{figure}[h]
\centerline{\psfig{figure=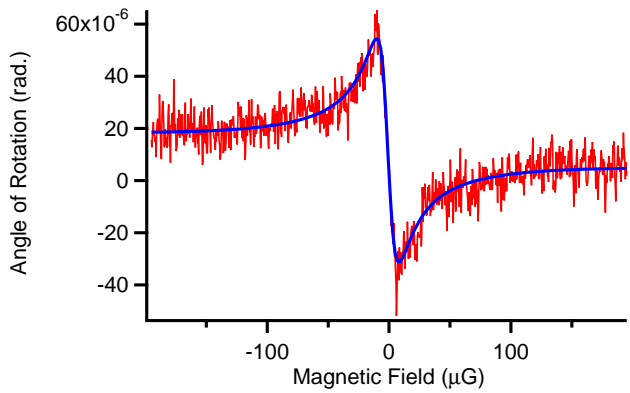,width=3.25in}}
\caption{Faraday rotation for the D1 transition.  The data were fit
to a dispersive function.  The width of the fit is
$12.6\,\mu\rm{G}$; the laser light power was $10.7\,\mu\rm{W}$.}
\label{D1Faraday}
\end{figure}

The intrinsic Zeeman relaxation rate for a given vapor cell can be
found from the low-light-power limit of the width of the Faraday
rotation curves such as the one shown in Fig.\ \ref{D1Faraday}
\cite{Bud98,Bud2005NIST}. This is done by fitting the amplitude as a
function of magnetic field to a dispersive lineshape to determine
the width at a given light power.  The intrinsic width is determined
by linear extrapolation of the power dependence of the widths.
Figure \ref{D1Faraday} shows typical data and fit for the D1
transition.

\begin{figure}[h]
\centerline{\psfig{figure=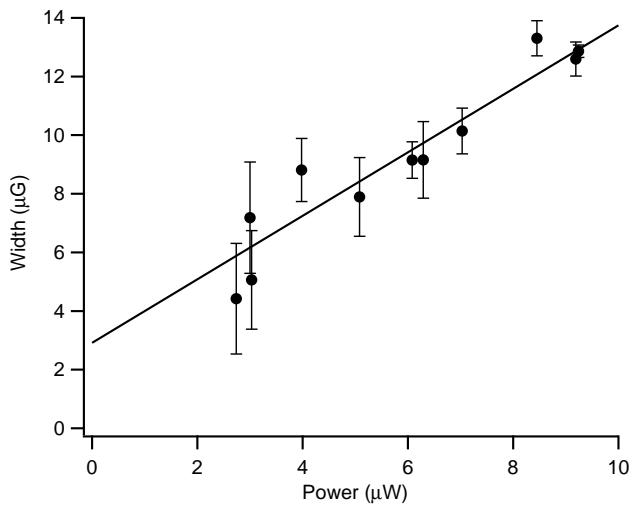,width=3.25in}}
\caption{Extrapolated NMOR widths for D1 transition. The intercept
${\rm B}=2.9(9)~{\rm \mu G}$ of the fit gives the intrinsic
relaxation rate. The error bars were increased to obtain a
$\chi_{\rm reduced}^2=1$.} \label{NMORWidths}
\end{figure}

Performing a linear extrapolation at light powers less than
$10\,\mu\rm{W}$ gives a light-independent magnetic field width of
$\Delta\rm{B}=2.9(9)~\mu {\rm G}$ for the data taken with the D1
transition (Fig. \ref{NMORWidths}).  Here $\Delta{\rm B}$ is the
difference in the magnetic field between the two extrema.  This
corresponds to a Zeeman relaxation rate of
\begin{align}
\gamma^{NMOR}/2\pi = g\mu \Delta {\rm B} = 2.0(6)\,{\rm
Hz}\label{Eq_gamma_result}
\end{align}
($g$ is the ground-state Land${\rm \acute{e}}$ factor; $\mu$ is the
Bohr magneton). The error bars for these data were increased by
about a factor of two compared to our estimates for individual fits
in order to obtain a reduced $\chi^2=1$.

Low-power data were also taken for the D2 transition.
In contrast to the D1 case, for the D2 transition evidence of
saturation was seen even at powers of just a few $\mu$W.
In this low-power regime, the signal sizes were on the order of the
background noise and we were unable to obtain satisfactory data for
the low powers required to get a linear fit (the D2 data alone would
indicate an intrinsic linewidth of $\Delta\rm{B}\lesssim 6~\mu {\rm
G}$).

It is interesting to compare the present results for the Zeeman
relaxation rate with the results of Ref.\ \cite{Bud2005NIST} for Rb
at similar vapor densities. In the work of Ref.\ \cite{Bud2005NIST}
the diameter of the vapor cells ranged from 3.4 cm to 10 cm and the
corresponding relaxation rates ranged between 3.5 and  $0.7\:{\rm
Hz}$. An analysis of possible relaxation mechanisms, including
spin-exchange relaxation, exchange of atoms with the metal reservoir
(the cell stem), etc., resulted in a conclusion that an additional
relaxation mechanism is present for Rb (and Cs \cite{Gra2005}),
possibly related to electron-spin randomization collisions on the
wall.

In the present experiment with a potassium cell, whose result is
given in Eq. \eqref{Eq_gamma_result}, the estimated spin-exchange
contribution to the relaxation rate is
\begin{align}
\frac{\gamma^{NMOR}_{\rm SE}}{2\pi}\approx
\frac{1}{2}\,n\,\sigma_{\rm SE}\, v_{\rm
rel}\prn{\frac{1}{2\pi}}\approx 0.4 \,{\rm Hz},\label{Eq:SE_NMOR}
\end{align}
where 1/2 is the appropriate ``nuclear slow-down factor'' for NMOR
in alkali atoms with nuclear spin $I=3/2$ \cite{Bud2005NIST}, $n$ is
the atomic density,
\begin{equation}
    v_{\rm rel} = \sqrt{8kT/\pi\mu_{red}}
\end{equation}
is the average relative speed of the atoms, $\mu_{red}$ is the
reduced mass of the colliding atoms, and $\sigma_{\rm SE}=2\cdot
10^{-14}\,{\rm cm^2}$ is the spin-exchange-collision cross section
\cite{Ale99_K,Ale99_K_ERRATUM}.

The relaxation rate due to exchange of atoms with the stem is
estimated to be $\approx 1\ $Hz with a large uncertainty due to the
difficulty of accounting for the specific geometry of the stem.
Therefore, the observed relaxation rate is not inconsistent with the
known sources of relaxation. Additional relaxation mechanisms that
were found with Rb and Cs contribute to the NMOR linewidth at a
level $\sim 1\ $Hz or less.
\section{Specific High-Power Feature of NMOR in K}
\begin{figure}[h]
\centerline{\psfig{figure=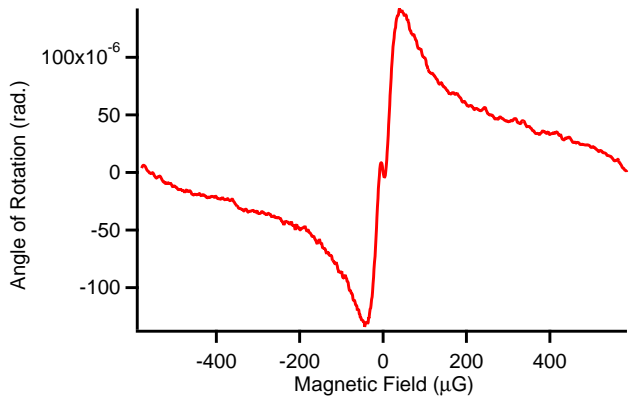,width=3.25in}}
\caption{Nested Faraday rotation feature for D1 signal at a light
power of $115\,{\rm \mu W}$. The origin of the narrow central
feature is interaction of the light with the hyperfine transitions
that are further away in frequency from where the laser is tuned
than the transitions contributing to the larger and broader feature.
These further-away transitions experience less power broadening. The
sign of an FM NMOR feature is determined by several factors,
including laser tuning with respect to a resonance, hyperfine
structure of the transitions, and the sign of the ground-state
g-factor (opposite for the two ground-state hyperfine components).}
\label{NestedFeature}
\end{figure}

At sufficiently high light power, a second narrower feature appears
imbedded within the main Faraday rotation curve (Fig.
\ref{NestedFeature}).

The nested Faraday curve is due to potassium's relatively small
ground-state hyperfine splitting, in comparison to its Doppler
width. The laser's central frequency is detuned $\approx 250 \:
\rm{MHz}$ to the low-frequency side of the absorption curve, so the
light preferentially interacts with the atoms in the $F=2$ ground
state. Thus, rotation of the light polarization is due to precession
of atomic alignment in the $F=2$ hyperfine level. However, there is
also a contribution from the $F=1$ hyperfine level since the
ground-state hyperfine structure is not well resolved.  As the laser
light power is increased, the more resonant transition from the
$F=2$ hyperfine level is power-broadened more than the less resonant
transition from the $F=1$ hyperfine level.  Thus, rotation due to
atomic alignment in the $F=1$ hyperfine level shows up as a narrower
feature nested within the power-broadened Faraday rotation due to
rotation of atomic alignment in the $F=2$ hyperfine level.
\section{Investigation of hyperfine-structure transitions}
Here we present the measurements of intrinsic linewidths and shifts
of the hyperfine-structure transitions performed with potassium
atoms in the paraffin-coated vapor cell.


Using the technique of optical pumping and radio-optical double
resonance \cite{Hap72}, we investigated the ``clock transition''
$F=1,M=0 \rightarrow F=2,M=0$ in $^{39}$K, which is to first order
insensitive to external DC magnetic fields and gradients.

The apparatus and procedure were similar to those described above,
except, instead of optical rotation, we measured the changes in
transmission of the light through optically pumped vapor when
radio-frequency magnetic field driving the hyperfine-structure
transitions was applied. This was done similarly to our earlier work
with Rb \cite{Bud2005NIST}. The highest signal amplitudes (for the
range of the experimental parameters used in this work) were
observed for the laser frequency detuned $\approx 250\ $MHz to the
low-frequency side of the peak of the D1 resonance.

Based on unsaturated resonance atomic absorption, in these
measurement, the atomic density was estimated as $8 \cdot 10^{9} \
\rm{cm}^{-3}$.

Radio-frequency transitions between hyperfine states of potassium
were driven using an HP8647A signal generator. The output of the
generator was connected via coaxial cable with a 9-mm diameter wire
loop terminating a coaxial cable, inserted inside the innermost
magnetic shielding next to the vapor cell. The signal generator was
referenced to the $10\ $MHz signal from a commercial
wireless-network-referenced atomic clock (Symmetricom TS2700); the
frequency of the generator's output was monitored with a Stanford
Research Systems SR620 frequency counter also referenced to the
TS2700 clock. The consistency of the frequency set by the generator
and the reading from the frequency counter was better than $0.1\
$Hz. Scans over the hyperfine transition were performed by sweeping
with 1-Hz steps the sine-wave frequency generated by the synthesizer
in the vicinity of the free-space separation of the ground-state
hyperfine components of $^{39}$K: 461,719,720.1 Hz \cite{Ari77}.

\begin{figure}[htbp]
    \centering
        \includegraphics[width=3.5 in]{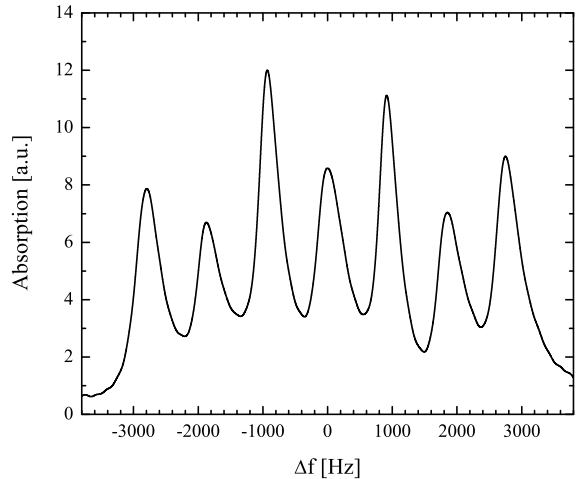}
    \caption{An example of the rf-spectrum recording showing resolved Zeeman components of the hyperfine-structure
    transition in a mG-sized field. The central peak corresponds to the 0-0 clock
transition. Light power: 30$\ \mu$W; rf power: 500 units used in
Fig. \ref{fig:zero}, and the resonances are strongly power
broadened.}
    \label{fig_Zeeman}
\end{figure}

The hyperfine-transition spectra were measured at fixed tuning of
the laser frequency and the longitudinal magnetic field of $\approx
1.4\ $mG, which was found to be sufficient to distinguish and
separate the 0-0 transition (an example of the recorded spectrum
showing resolved Zeeman components is shown in Fig.
\ref{fig_Zeeman}).

Data were taken for a range of light and rf powers. Averaging over
$\approx 50-200$ scans was necessary because of weak amplitudes of
rf resonances at low powers. The duration of a single scan was
$\approx 1/2\ $min dominated by the settling time of the rf
frequency synthesizer (0.2 s).

\begin{figure}[htbp]
    \centering
        \includegraphics[width=3.5 in]{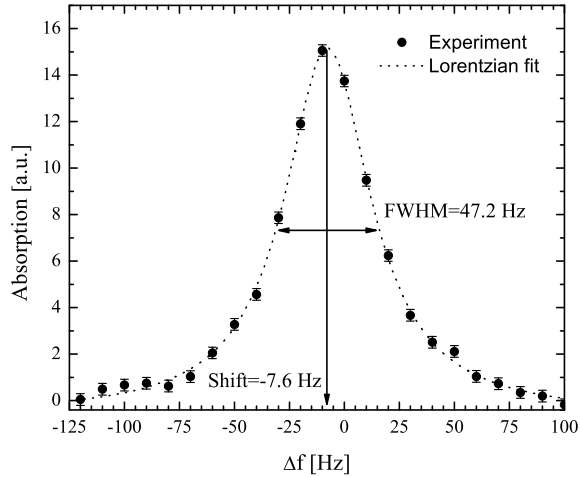}
        \caption{Example of a measured spectrum of the isolated 0-0 transition. Light power:  $12\
        \mu$W; rf power: 35 units used in Fig. \ref{fig:zero}. The plot represents an average of approximately 50 scans.
        The fit Lorentzian linewidth is $47.2\ $Hz (FWHM), larger than
        the ``intrinsic'' width of $10.6\ $Hz due to residual light- and rf-power broadening.
        }
    \label{fig:example}
\end{figure}

\begin{figure}[h!tbp]
    \centering
        \includegraphics[width=3.5 in]{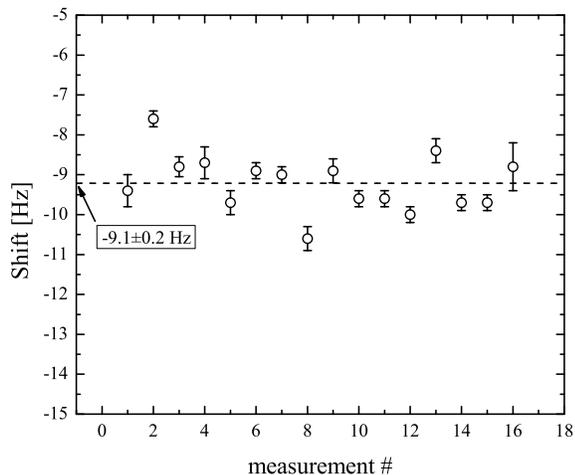}
        \caption{Experimental results for the shift of the frequency of the 0-0 clock transition from its
        free-space value. The plot shows all the results obtained with different light and rf powers;
        no systematic dependence on these parameters was observed. Negative shift means that,
        in the paraffin-coated cell, the hyperfine transition frequency is lower than that for an atom in free space.
        Error bars represent the uncertainty obtained from an individual fit. The spread of the points exceeds what would
        be expected from the size of these error bars, presumably due to slow drifts of experimental parameters (such as cell temperature)
        that may result in apparent shift of the line.}
    \label{fig:shift}
\end{figure}

\begin{figure}[h!tbp]
    \centering
        \includegraphics[width=3.5 in]{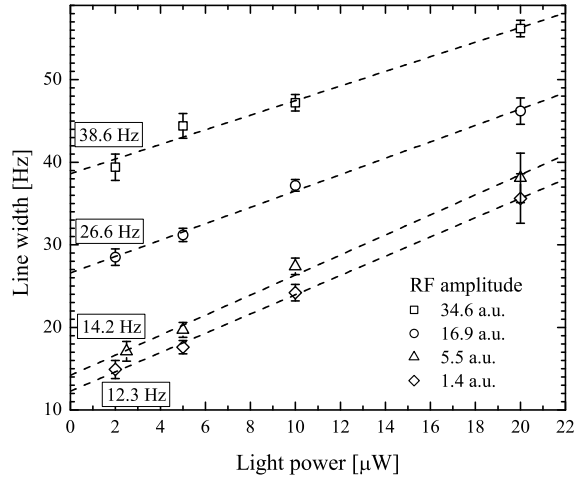}
        \caption{Linewidths measured for different laser- and rf-powers. The fit lines correspond
        to performed extrapolation of the linewidth to zero in the light power. The zero-power widths were
        used to perform extrapolation in rf power, see Fig. \ref{fig:zero}.}
    \label{fig:widths}
\end{figure}

\begin{figure}[h!tbp]
    \centering
        \includegraphics[width=3.5 in]{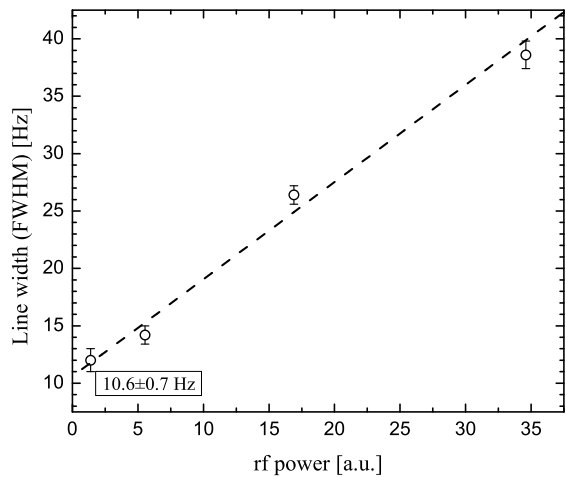}
        \caption{Extrapolation to zero rf power of the zero-light-power values of the rf-transition width (see Fig. \ref{fig:widths}).}
    \label{fig:zero}
\end{figure}

An example of an experimental scan over the isolated 0-0 transition
is shown in Fig. \ref{fig:example}.
In the case of K, the wavelength of the hyperfine-structure
transition $\lambda \approx 65$ cm is significantly larger than the
vapor cell, meaning that the system is in the Dicke-narrowing
regime, and there is no Doppler pedestal in the resonance spectra.
Hyperfine frequency shifts and relaxation rate were found from the
low-light and low-rf-power limit of the double-resonance curves. We
fit data for a given rf and light power with a Lorentz profile and
extract the width and the central frequency of the resonance. The
linewidths dependence on the rf and light power due to power
broadening, in the low-power limit, becomes linear and it is
possible to perform double extrapolation of the resonance center and
linewidths to zero power, and thus to determine the intrinsic
hyperfine-transition frequency shift and relaxation rate for a given
cell. Results of the fitting for each light and rf power are shown
in Figs. \ref{fig:shift} and \ref{fig:widths}. Extrapolation to zero
power was first performed for light power [Fig. \ref{fig:widths}]
and then for rf power [Fig. \ref{fig:zero}]. Within experimental
uncertainties, the central position of the resonance did not exhibit
any systematic dependence on the light and rf power.

Finally, based on all measurement, the shift was found to be
$-9.1(2)\ $Hz, and the intrinsic linewidth was found to be $10.6(7)\
$Hz.

\section{Interpretation of the hyperfine-transition linewidth and shift}
Here we present the analysis of the above results that follows a
similar analysis that was carried out for Rb in Ref.
\cite{Bud2005NIST}.

The spin-exchange-relaxation contribution to the linewidth of the
hyperfine resonance can be estimated as follows. For a single
isotope with nuclear spin $I$ present in the cell, the spin-exchange
contribution is given by an equation [similar to Eq.
\eqref{Eq:SE_NMOR}]:
\begin{equation}
    \frac{\gamma^{hf}_{\rm SE}}{2\pi} = \frac{R(I)nv_{\rm rel}\sigma_{\rm SE}}{\pi},
\end{equation}
where $R(I)=(6I+1)/(8I+4)$ is the nuclear slow-down factor for the
0-0 transition. With the temperature and density used in the present
experiment, one calculates:
\begin{equation}
    \frac{\gamma^{hf}_{\rm SE}}{2\pi} \approx 2.2 \ {\rm Hz}.\label{Eq:gamma_h_SE}
\end{equation}

The relaxation rate due to exchange of atoms with the stem must be
consistent with that for the case of NMOR. With the adopted
definitions of $\gamma^{NMOR}$, and $\gamma^{hf}$, we estimate
(again, with a large uncertainty):
\begin{equation}
    \frac{\gamma^{hf}_{\rm stem}}{2\pi} \approx 2\ {\rm Hz}.\label{Eq:gamma_h_stem}
\end{equation}

Adiabatic wall collisions are the most ``gentle'' collisions of
alkali atoms with the wall coating. They do not result in population
transfer or Zeeman decoherence, but they introduce a phase-shift
$\phi$ between hyperfine states. The phase-shift acquired by the
atoms have a statistical character, which leads to a shift of the
resonance and contributes to its width:
\begin{equation}
    \frac{\gamma^{hf}_{a}}{2\pi} = \frac{\phi^2}{\pi t_c},
\end{equation}
where $t_c = 4R/3 \bar{\upsilon}$ is the characteristic time between
wall collisions, $R$ is the cell radius and $\bar{\upsilon}$ is the
average speed of atoms. The average phase-shift $\phi$ can be
determined from the fact that the resonance frequency shift is
$\delta \nu = \phi/(2\pi t_c)$. This yields
\begin{equation}
    \phi \approx 2.8\cdot 10 ^{-3},
\end{equation}
which is in the range that may be expected from the phase shifts
observed with the two Rb isotopes and Cs (see Refs.
\cite{Bud2005NIST,VanierAudoin} and references therein) if one
assumes linear scaling of the phase shift with the hyperfine
interval. (The phase shift is also a strong function of the surface
binding energy. The observed phase-shift scaling suggests that the
surface binding energies are comparable for Rb, Cs, and K.)

The contribution to the hyperfine transition linewidth calculated on
this basis is:
\begin{equation}
    \frac{\gamma^{hf}_{a}}{2\pi} \approx 0.05\ \rm{Hz}. \label{Eq:gamma_h_a}
\end{equation}

Summing the contributions from Eqs.
\eqref{Eq:gamma_h_stem},\eqref{Eq:gamma_h_SE}, and
\eqref{Eq:gamma_h_a}, we can account for about $4-5\ $Hz out of the
total observed hyperfine-transition width of $10.6(7)\ $Hz.

A possible additional contribution to relaxation could be coming
from alkali-vapor density dependent spin-randomization wall
collisions, as discussed for Rb and Cs in Refs.
\cite{Bud2005NIST,Gra2005}. In this experiment, this contribution
may have been more pronounced for the hyperfine transitions because
they were measured at higher potassium vapor density. In future work
this effect could be explored more definitively by performing Zeeman
and hyperfine relaxation measurements concurrently, and as a
function of the potassium vapor density. It would also be
interesting to measure the rate of longitudinal (T1) relaxation on
the 0-0 transition to which adiabatic collisions do not contribute.

%

\section{Conclusion}
We have studied nonlinear magneto-optical rotation and Zeeman
relaxation rates of potassium contained in a 3-cm-diameter
paraffin-coated cell. The magnitude of the nonlinear rotation due to
atomic alignment is significantly suppressed in K compared to Rb and
Cs owing to the small upper-state hyperfine-structure intervals. The
fact that the ground-state hyperfine separation is comparable to the
Doppler width leads to the appearance of a peculiar nested feature
observed in the magnetic-field dependence of the optical rotation at
relatively high light powers. The measured intrinsic ground-state
relaxation rate of about 2 Hz is consistent with estimated
relaxation due to spin-exchange collisions and the exchange of atoms
between the volume of the vapor cell and the stem. This is evidence
of the high quality of the coating and a direct illustration of the
remarkable properties of paraffin as the coating material
\cite{Rob58,BouBro66}: it takes on the order of $10^4$ collisions of
a polarized atom with the paraffin-coated wall to induce relaxation.
We have also measured the frequency shift and linewidth of the
hyperfine 0-0 clock transition. The measured frequency shift in
$^{39}$K is smaller than that in Rb and Cs, approximately in
proportion to the transition frequency. This indicates that the
contribution of the adiabatic collisions to the hyperfine-transition
linewidth is very small [Eq. \eqref{Eq:gamma_h_a}]. This supports
the idea that if adiabatic relaxation was the dominant relaxation
mechanism, it would be beneficial to use potassium (rather than the
higher transition frequency alkalis such as $^{87}$Rb and Cs) for
secondary frequency references in coated cells. Unfortunately,
additional relaxation mechanisms exist (as seen in the present
measurement), which largely remove the possible advantages.

\section{Acknowledgements}
The authors are grateful to E. B. Alexandrov, D. F. Jackson Kimball,
D. English, M. Auzinsh, S. M. Rochester, J. Higbie, and W. Gawlik
for helpful discussions. This work was supported by DOD MURI grant
N-00014-05-1-0406, and by NSF US-Poland collaboration grant and REU
supplement.
\bibliography{NMOBIB}

\begin{thebibliography}{18}
\expandafter\ifx\csname natexlab\endcsname\relax\def\natexlab#1{#1}\fi
\expandafter\ifx\csname bibnamefont\endcsname\relax
  \def\bibnamefont#1{#1}\fi
\expandafter\ifx\csname bibfnamefont\endcsname\relax
  \def\bibfnamefont#1{#1}\fi
\expandafter\ifx\csname citenamefont\endcsname\relax
  \def\citenamefont#1{#1}\fi
\expandafter\ifx\csname url\endcsname\relax
  \def\url#1{\texttt{#1}}\fi
\expandafter\ifx\csname urlprefix\endcsname\relax\def\urlprefix{URL }\fi
\providecommand{\bibinfo}[2]{#2}
\providecommand{\eprint}[2][]{\url{#2}}

\bibitem[{\citenamefont{Budker et~al.}(2005)\citenamefont{Budker, Hollberg,
  Kimball, Kitching, Pustelny, and Yashchuk}}]{Bud2005NIST}
\bibinfo{author}{\bibfnamefont{D.}~\bibnamefont{Budker}},
  \bibinfo{author}{\bibfnamefont{L.}~\bibnamefont{Hollberg}},
  \bibinfo{author}{\bibfnamefont{D.~F.} \bibnamefont{Kimball}},
  \bibinfo{author}{\bibfnamefont{J.}~\bibnamefont{Kitching}},
  \bibinfo{author}{\bibfnamefont{S.}~\bibnamefont{Pustelny}}, \bibnamefont{and}
  \bibinfo{author}{\bibfnamefont{V.~V.} \bibnamefont{Yashchuk}},
  \bibinfo{journal}{Physical Review A} \textbf{\bibinfo{volume}{71}},
  \bibinfo{pages}{012903} (\bibinfo{year}{2005}).

\bibitem[{\citenamefont{Balabas et~al.}(2006)\citenamefont{Balabas, Budker,
  Kitching, Schwindt, and Stalnaker}}]{Bal2006}
\bibinfo{author}{\bibfnamefont{M.~V.} \bibnamefont{Balabas}},
  \bibinfo{author}{\bibfnamefont{D.}~\bibnamefont{Budker}},
  \bibinfo{author}{\bibfnamefont{J.}~\bibnamefont{Kitching}},
  \bibinfo{author}{\bibfnamefont{P.~D.~D.} \bibnamefont{Schwindt}},
  \bibnamefont{and} \bibinfo{author}{\bibfnamefont{J.~E.}
  \bibnamefont{Stalnaker}}, \bibinfo{journal}{Journal of the Optical Society of
  America B (Optical Physics)} \textbf{\bibinfo{volume}{23}}
  (\bibinfo{year}{2006}).

\bibitem[{\citenamefont{Graf et~al.}(2005)\citenamefont{Graf, Kimball,
  Rochester, Kerner, Wong, Budker, Alexandrov, Balabas, and
  Yashchuk}}]{Gra2005}
\bibinfo{author}{\bibfnamefont{M.~T.} \bibnamefont{Graf}},
  \bibinfo{author}{\bibfnamefont{D.~F.} \bibnamefont{Kimball}},
  \bibinfo{author}{\bibfnamefont{S.~M.} \bibnamefont{Rochester}},
  \bibinfo{author}{\bibfnamefont{K.}~\bibnamefont{Kerner}},
  \bibinfo{author}{\bibfnamefont{C.}~\bibnamefont{Wong}},
  \bibinfo{author}{\bibfnamefont{D.}~\bibnamefont{Budker}},
  \bibinfo{author}{\bibfnamefont{E.~B.} \bibnamefont{Alexandrov}},
  \bibinfo{author}{\bibfnamefont{M.~V.} \bibnamefont{Balabas}},
  \bibnamefont{and} \bibinfo{author}{\bibfnamefont{V.~V.}
  \bibnamefont{Yashchuk}}, \bibinfo{journal}{Physical Review A}
  \textbf{\bibinfo{volume}{72}}, \bibinfo{pages}{023401}
  (\bibinfo{year}{2005}).

\bibitem[{\citenamefont{Yashchuk et~al.}(2004)\citenamefont{Yashchuk, Granwehr,
  Kimball, Rochester, Trabesinger, Urban, Budker, and Pines}}]{Yas2004}
\bibinfo{author}{\bibfnamefont{V.~V.} \bibnamefont{Yashchuk}},
  \bibinfo{author}{\bibfnamefont{J.}~\bibnamefont{Granwehr}},
  \bibinfo{author}{\bibfnamefont{D.~F.} \bibnamefont{Kimball}},
  \bibinfo{author}{\bibfnamefont{S.~M.} \bibnamefont{Rochester}},
  \bibinfo{author}{\bibfnamefont{A.~H.} \bibnamefont{Trabesinger}},
  \bibinfo{author}{\bibfnamefont{J.~T.} \bibnamefont{Urban}},
  \bibinfo{author}{\bibfnamefont{D.}~\bibnamefont{Budker}}, \bibnamefont{and}
  \bibinfo{author}{\bibfnamefont{A.}~\bibnamefont{Pines}},
  \bibinfo{journal}{Physical Review Letters} \textbf{\bibinfo{volume}{93}},
  \bibinfo{pages}{160801} (\bibinfo{year}{2004}).

\bibitem[{\citenamefont{Xu et~al.}(2006)\citenamefont{Xu, Yashchuk, Donaldson,
  Rochester, Budker, and Pines}}]{Xu2006IMAG}
\bibinfo{author}{\bibfnamefont{S.}~\bibnamefont{Xu}},
  \bibinfo{author}{\bibfnamefont{V.~V.} \bibnamefont{Yashchuk}},
  \bibinfo{author}{\bibfnamefont{M.~H.} \bibnamefont{Donaldson}},
  \bibinfo{author}{\bibfnamefont{S.~M.} \bibnamefont{Rochester}},
  \bibinfo{author}{\bibfnamefont{D.}~\bibnamefont{Budker}}, \bibnamefont{and}
  \bibinfo{author}{\bibfnamefont{A.}~\bibnamefont{Pines}},
  \bibinfo{journal}{Submitted}  (\bibinfo{year}{2006}).

\bibitem[{\citenamefont{Aleksandrov et~al.}(1999)\citenamefont{Aleksandrov,
  Balabas, Vershovskii, Okunevich, and Yakobson}}]{Ale99_K}
\bibinfo{author}{\bibfnamefont{E.~B.} \bibnamefont{Aleksandrov}},
  \bibinfo{author}{\bibfnamefont{M.~V.} \bibnamefont{Balabas}},
  \bibinfo{author}{\bibfnamefont{A.~K.} \bibnamefont{Vershovskii}},
  \bibinfo{author}{\bibfnamefont{A.~I.} \bibnamefont{Okunevich}},
  \bibnamefont{and} \bibinfo{author}{\bibfnamefont{N.~N.}
  \bibnamefont{Yakobson}}, \bibinfo{journal}{Optics and Spectroscopy}
  \textbf{\bibinfo{volume}{87}}, \bibinfo{pages}{329} (\bibinfo{year}{1999}).

\bibitem[{\citenamefont{Aleksandrov et~al.}(2002)\citenamefont{Aleksandrov,
  Balabas, Vershovskii, Okunevich, and Yakobson}}]{Ale99_K_ERRATUM}
\bibinfo{author}{\bibfnamefont{E.~B.} \bibnamefont{Aleksandrov}},
  \bibinfo{author}{\bibfnamefont{M.~V.} \bibnamefont{Balabas}},
  \bibinfo{author}{\bibfnamefont{A.~K.} \bibnamefont{Vershovskii}},
  \bibinfo{author}{\bibfnamefont{A.~I.} \bibnamefont{Okunevich}},
  \bibnamefont{and} \bibinfo{author}{\bibfnamefont{N.~N.}
  \bibnamefont{Yakobson}}, \bibinfo{journal}{Optics and Spectroscopy}
  \textbf{\bibinfo{volume}{93}}, \bibinfo{pages}{488(E)}
  (\bibinfo{year}{2002}).

\bibitem[{\citenamefont{Budker et~al.}(2002{\natexlab{a}})\citenamefont{Budker,
  Gawlik, Kimball, Rochester, Yashchuk, and Weis}}]{Bud2002RMP}
\bibinfo{author}{\bibfnamefont{D.}~\bibnamefont{Budker}},
  \bibinfo{author}{\bibfnamefont{W.}~\bibnamefont{Gawlik}},
  \bibinfo{author}{\bibfnamefont{D.~F.} \bibnamefont{Kimball}},
  \bibinfo{author}{\bibfnamefont{S.~M.} \bibnamefont{Rochester}},
  \bibinfo{author}{\bibfnamefont{V.~V.} \bibnamefont{Yashchuk}},
  \bibnamefont{and} \bibinfo{author}{\bibfnamefont{A.}~\bibnamefont{Weis}},
  \bibinfo{journal}{Reviews of Modern Physics} \textbf{\bibinfo{volume}{74}},
  \bibinfo{pages}{1153} (\bibinfo{year}{2002}{\natexlab{a}}).

\bibitem[{\citenamefont{Budker et~al.}(2000)\citenamefont{Budker, Kimball,
  Rochester, and Yashchuk}}]{Bud2000AOC}
\bibinfo{author}{\bibfnamefont{D.}~\bibnamefont{Budker}},
  \bibinfo{author}{\bibfnamefont{D.~F.} \bibnamefont{Kimball}},
  \bibinfo{author}{\bibfnamefont{S.~M.} \bibnamefont{Rochester}},
  \bibnamefont{and} \bibinfo{author}{\bibfnamefont{V.~V.}
  \bibnamefont{Yashchuk}}, \bibinfo{journal}{Physical Review Letters}
  \textbf{\bibinfo{volume}{85}}, \bibinfo{pages}{2088} (\bibinfo{year}{2000}).

\bibitem[{\citenamefont{Budker et~al.}(2002{\natexlab{b}})\citenamefont{Budker,
  Kimball, Yashchuk, and Zolotorev}}]{Bud2002FM}
\bibinfo{author}{\bibfnamefont{D.}~\bibnamefont{Budker}},
  \bibinfo{author}{\bibfnamefont{D.~F.} \bibnamefont{Kimball}},
  \bibinfo{author}{\bibfnamefont{V.~V.} \bibnamefont{Yashchuk}},
  \bibnamefont{and}
  \bibinfo{author}{\bibfnamefont{M.}~\bibnamefont{Zolotorev}},
  \bibinfo{journal}{Physical Review A (Atomic, Molecular, and Optical Physics)}
  \textbf{\bibinfo{volume}{65}}, \bibinfo{pages}{055403}
  (\bibinfo{year}{2002}{\natexlab{b}}).

\bibitem[{\citenamefont{Alexandrov et~al.}(2002)\citenamefont{Alexandrov,
  Balabas, Budker, English, Kimball, Li, and Yashchuk}}]{AleLIAD}
\bibinfo{author}{\bibfnamefont{E.~B.} \bibnamefont{Alexandrov}},
  \bibinfo{author}{\bibfnamefont{M.~V.} \bibnamefont{Balabas}},
  \bibinfo{author}{\bibfnamefont{D.}~\bibnamefont{Budker}},
  \bibinfo{author}{\bibfnamefont{D.}~\bibnamefont{English}},
  \bibinfo{author}{\bibfnamefont{D.~F.} \bibnamefont{Kimball}},
  \bibinfo{author}{\bibfnamefont{C.~H.} \bibnamefont{Li}}, \bibnamefont{and}
  \bibinfo{author}{\bibfnamefont{V.~V.} \bibnamefont{Yashchuk}},
  \bibinfo{journal}{Physical Review A} \textbf{\bibinfo{volume}{66}},
  \bibinfo{pages}{042903/1} (\bibinfo{year}{2002}).

\bibitem[{\citenamefont{Budker et~al.}(1998)\citenamefont{Budker, Yashchuk, and
  Zolotorev}}]{Bud98}
\bibinfo{author}{\bibfnamefont{D.}~\bibnamefont{Budker}},
  \bibinfo{author}{\bibfnamefont{V.}~\bibnamefont{Yashchuk}}, \bibnamefont{and}
  \bibinfo{author}{\bibfnamefont{M.}~\bibnamefont{Zolotorev}},
  \bibinfo{journal}{Physical Review Letters} \textbf{\bibinfo{volume}{81}},
  \bibinfo{pages}{5788} (\bibinfo{year}{1998}).

\bibitem[{\citenamefont{Vanier et~al.}(1974)\citenamefont{Vanier, Simard, and
  Boulanger}}]{Van74}
\bibinfo{author}{\bibfnamefont{J.}~\bibnamefont{Vanier}},
  \bibinfo{author}{\bibfnamefont{J.~F.} \bibnamefont{Simard}},
  \bibnamefont{and} \bibinfo{author}{\bibfnamefont{J.~S.}
  \bibnamefont{Boulanger}}, \bibinfo{journal}{Physical Review A (General
  Physics)} \textbf{\bibinfo{volume}{9}}, \bibinfo{pages}{1031}
  (\bibinfo{year}{1974}).

\bibitem[{\citenamefont{Vanier and Audoin}(1989)}]{VanierAudoin}
\bibinfo{author}{\bibfnamefont{J.}~\bibnamefont{Vanier}} \bibnamefont{and}
  \bibinfo{author}{\bibfnamefont{C.}~\bibnamefont{Audoin}},
  \emph{\bibinfo{title}{The quantum physics of atomic frequency standards}}
  (\bibinfo{publisher}{A. Hilger}, \bibinfo{address}{Bristol ; Philadelphia},
  \bibinfo{year}{1989}).

\bibitem[{\citenamefont{Happer}(1972)}]{Hap72}
\bibinfo{author}{\bibfnamefont{W.}~\bibnamefont{Happer}},
  \bibinfo{journal}{Reviews of Modern Physics} \textbf{\bibinfo{volume}{44}},
  \bibinfo{pages}{169} (\bibinfo{year}{1972}).

\bibitem[{\citenamefont{Arimondo et~al.}(1977)\citenamefont{Arimondo, Inguscio,
  and Violino}}]{Ari77}
\bibinfo{author}{\bibfnamefont{E.}~\bibnamefont{Arimondo}},
  \bibinfo{author}{\bibfnamefont{M.}~\bibnamefont{Inguscio}}, \bibnamefont{and}
  \bibinfo{author}{\bibfnamefont{P.}~\bibnamefont{Violino}},
  \bibinfo{journal}{Reviews of Modern Physics} \textbf{\bibinfo{volume}{49}},
  \bibinfo{pages}{31} (\bibinfo{year}{1977}).

\bibitem[{\citenamefont{Robinson et~al.}(1958)\citenamefont{Robinson, Ensberg,
  and Dehmelt}}]{Rob58}
\bibinfo{author}{\bibfnamefont{H.}~\bibnamefont{Robinson}},
  \bibinfo{author}{\bibfnamefont{E.}~\bibnamefont{Ensberg}}, \bibnamefont{and}
  \bibinfo{author}{\bibfnamefont{H.}~\bibnamefont{Dehmelt}},
  \bibinfo{journal}{Bulletin of the American Physical Society}
  \textbf{\bibinfo{volume}{3}}, \bibinfo{pages}{9} (\bibinfo{year}{1958}).

\bibitem[{\citenamefont{Bouchiat and Brossel}(1966)}]{BouBro66}
\bibinfo{author}{\bibfnamefont{M.~A.} \bibnamefont{Bouchiat}} \bibnamefont{and}
  \bibinfo{author}{\bibfnamefont{J.}~\bibnamefont{Brossel}},
  \bibinfo{journal}{Physical Review} \textbf{\bibinfo{volume}{147}},
  \bibinfo{pages}{41} (\bibinfo{year}{1966}).

\end{thebibliography}

\end{document}